# Title: High-temperature Phonon Coherence and Tunneling Effect in Semiconductor Superlattices

**Authors:** Zhi-Ming Geng[1†], Jin-Shan Yao[1†], Ying-Bin Cheng[1†], Xue-Jun Yan[1,3†], Jian Zhou[1,2], En-Rui Zhang[1], Jia-Yi Li[1], Ming-Qian Yuan[1], Xing Fan[1], Yu Deng[1,2,3], Hong Lu[1,2,3]*, Ming-Hui Lu[1,2,3]*, Yan-Feng Chen[1]*

**Affiliations:**

[1]National Laboratory of Solid State Microstructures & Department of Materials Science and Engineering, College of Engineering and Applied Sciences, Nanjing University; Nanjing, 210093, China.

[2]Collaborative Innovation Center of Advanced Microstructures, Nanjing University; Nanjing, 210093, China.

[3]Jiangsu Key Laboratory of Artificial Functional Materials, Nanjing University; Nanjing, 210093, China.

*Corresponding author. Email: hlu@nju.edu.cn, luminghui@nju.edu.cn, yfchen@nju.edu.cn

†These authors contributed equally to this work.

**Abstract:** Phonons, the quanta of lattice vibrations, are primary heat carriers for semiconductors and dielectrics. The demand of effective phonon manipulation urgently emerges, because the thermal management is crucial for the ongoing development of micro/nano semiconductor devices towards higher integration and power densities[1, 2]. Phonons also show wave-particle duality, while they are commonly treated as particle flows in current semiconductor structures[3, 4]. However, it sees constraints when the structure size reduces to nano and atomic scales, where the wave behavior of phonons begins to dominate, and studies of these phonon behaviors and their manipulations become long-standing challenges in experiments[5]. Here we show the experimental realization of coherent phonon transport, a wave-based thermal conduction fashion, in semiconductor structures. We report the successful observation of robust phonon coherence and tunneling effect in InAs/AlAs superlattices over an extensive temperature range up to 500 K, a breakthrough towards practical-application temperature for semiconductors compared with cryogenic conditions[6]. Our results demonstrate that the phonon coherence is robust even at a record-high interface density due to the dominating long-wavelength phonons, and the first-principles calculations clearly reveal their wave-particle duality. This revelation heralds a promising pathway towards efficient thermal phonon engineering at extreme scales, holding implications for a broad spectrum of semiconductor device applications, including microelectronics, optoelectronics, and thermoelectrics.

**Main Text:**

Heat conduction in solid matter mainly depends on two categories of energy carriers: phonons and electrons. Phonons, defined as the quanta of lattice vibrations, are the major heat carriers for semiconductors and insulators. It is of great significance to study phonon transport in nanostructured semiconductors, as heat dissipation has become a critical issue in very large scale integration driven by Moore's Law[7]. The demand is around the corner, which calls for more effective thermal management technologies with extended capabilities on nano and even atomic-scale structures. More in-depth insights into phonon transport mechanisms in such structures are thus required[4, 8]. Like other quantum particles, phonons also show the wave-particle duality, while they are treated as particle flows in most cases[3]. As the structure scales are being pushed to minimum limit the wave nature of phonons must be considered, and the thermal transport modulation may take the advantage of it. In the wave regime, phonons can traverse through defects, interfaces, and boundaries, with their phase information preserved, presenting a coherent phonon transport. As one-dimensional phononic crystals, superlattices offer an ideal platform to observe coherent phonons by enabling possible phonon wave interference[5]. Observations of coherent phonons had been reported by different approaches[10-13]. However, it is more challenging to experimentally observe such behavior in thermal transport since it is an integrated effect of full-spectrum phonons. Therefore, collective phonon behaviors have drawn considerable research interests due to their direct contributions to heat conduction[14, 15], and also the promise in its manipulation[16-18].

Coherent phonon transport in superlattices had been investigated theoretically[19-21]. It is found that there exists a minimum in cross-plane thermal conductivity as a function of period length[22-24]. In addition, the cross-plane thermal conductivity shows a specific period-dependent fashion that it increases with decreasing period length, which is derived from phonon tunneling[25]. The minimum thermal conductivity has been considered as an indicator of phonons transport transition from incoherent to coherent[22]. However, the experimental observation of the coherent phonon transport still remains an extreme challenge due to the imperfections of materials, which are unavoidable[26-28], such as defects[29, 30], poor interfacial quality[31], disordered periodicity, etc[32-35]. To date, only a few reports have demonstrated the evidence of coherent phonon transport in superlattices. For example, Ravichandran et al.[36] chose $SrTiO_3$-based superlattices for their relatively long phonon coherence length[37-39] and decent Umklapp peak, to evidently observe the coherent phonons at room temperature. Luckyanova et al.[6] reported a pioneering work on GaAs/AlAs superlattices and showed the dependence of thermal conductivity on the period numbers. The linear dependence indicated the presence of coherent phonons. However, this result is considered a piece of indirect evidence, and more conclusive evidence, i.e., the period-dependent fashion of thermal conductivity or phonon tunneling is still absent in III-V binary superlattices. Part of the reason is that the coherence length of GaAs is much shorter, in the order of 1-2 nm[36, 38]. In addition, the coherent phonon transport in the GaAs/AlAs superlattices is limited to low temperatures, up to 150 K.

In this work, we employ a strategy to construct a superlattice system with large mass contrast so that the long-wavelength phonons can be promoted and dominate the thermal transport[40]. The coherent phonon transport and tunneling effect over a broad temperature range from 100 to 500 K are experimentally demonstrated in a series of strain-compensated InAs/AlAs superlattices. The robust phonon coherence is facilitated by the high quality superlattice structures and record-high interface density (atomic single-layer). The first-principles calculations show excellent agreement with the experiments, demonstrating the dominant coherent transport mechanism and the wave-particle duality of phonons. This work offers an important reference for thermal manipulation in an array of nanostructures, and potential band engineering in phononic crystals[17].

**Experimental demonstration of robust phonon coherence and tunneling effect**

InAs/AlAs superlattices proposed herein were designed to feature different period lengths, and accordingly, different interface densities. The period length was tuned by constructing different numbers of InAs (AlAs) monolayers (ML), for which the superlattices can be named $InAs_{nML}/AlAs_{nML}$, where n refers to the number of InAs (AlAs) MLs. One ML contains one layer of In (Al) atoms and one layer of As atoms. In this zincblende structure, one ML corresponds to half the size of the lattice constant. In this study, the number of ML n is between 1 to 5, so that the period length is below the critical thickness in this strain-compensated superlattice[41]. To study the dependence of thermal conductivity on the period length, the thicknesses of these superlattices were kept the same, at 250 nm. This thickness is larger than the heat penetration depth, so the superlattice can be treated as an infinite model, and hence the size effects can be ruled out[42]. Two additional series of 1/1 and 2/2 ML superlattices with different period numbers (i.e. different total thicknesses) were grown to further investigate possible coherent phonon behavior. The superlattice samples were grown by using molecular beam epitaxy (MBE) on InP substrates. The periodic structure and the period length are confirmed by X-ray diffraction (XRD) and scanning transmission electron microscope (STEM). The cross-plane thermal conductivities were measured using the time-domain thermoreflectance (TDTR) technique (Extended Data Fig.1). Experimental details are included in the Methods section.

The principal findings of this work are given in Fig. 1. The period dependence of the thermal conductivity for InAs/AlAs superlattices at room temperature is plotted in Fig. 1a (representative data fitting for the TDTR measurements is given in Extended Data Fig. 2). The thermal conductivity increases evidently with decreasing period length (or increasing interface density) for the relatively-short period (1/1, 2/2 and 3/3 ML) superlattices, in which the period lengths are comparable to the phonon coherence length[36, 38]. This specific trend reveals coherent phonon transport dominates the thermal transport across the superlattice interfaces and the phonons exhibit wave-like behavior as illustrated in the insets. Otherwise, if phonons show particle-like behavior, then the thermal conductivity would exhibit an opposite trend because the particles should be more efficiently scattered by the increased interface density. More specifically, for the superlattices with extremely short periods, the trend found in the short-period InAs/AlAs superlattices is considered to be derived from phonon tunneling, for which the critical period length was demonstrated as just a few nanometers[25]. For the InAs/AlAs superlattices with relatively long periods (i. e. 4/4 and 5/5 ML), the thermal conductivity remains almost unchanged, yielding the minimum in these superlattices. This scenario indicates an increasing scattering effect and then a competition mechanism between the wave- and particle-like behaviors of phonons[36]. The scattering effect herein can be understood as interface scattering in the particle regime or Bragg scattering in the wave regime[31].

We further measured the thermal conductivity of InAs/AlAs superlattices at various temperatures to explore the robustness of the phonon coherence. As shown in Fig. 1b, in a broad temperature range from 100 to 500 K, the thermal conductivity maintains this period-dependent trend that increases with decreasing period length. This is a piece of strong evidence indicating that the coherent phonon transport behavior in the InAs/AlAs superlattices is robust even at a relatively high temperature of up to 500 K. It is easier to realize coherent transport for phonons at lower temperatures because the long-wavelength phonons are more dominant in thermal transport[6, 36]. In detail, such a period-dependent trend of thermal conductivity becomes more "linear" at lower temperatures, which is especially pronounced for the data at 100 K. This phenomenon shows that in the 4/4 ML superlattice, the contribution of coherent phonons to thermal transport becomes larger at low temperatures so that the thermal conductivity becomes higher than that of the 5/5 ML sample. (The curves shown in Fig. 1b are offset to better compare the period-dependent trends. The original data can be found in Extended Data Fig. 3.)

We further investigate the relationship between the thermal conductivities and the period numbers (i.e. total thicknesses) of the short-period superlattices (1/1 and 2/2 ML) at room temperatures, which is shown in Fig .1C. It is found that the thermal conductivities of the two superlattices both show linear dependence upon the increasing period numbers, indicating the phonon transport is primarily coherent in these superlattices. A similar trend had been reported previously in another III-V system, AlAs/GaAs superlattices[6]. Such linearity of the thermal conductivity versus period number can be considered a potential indicator of phonon coherence in superlattices. Although phonon is a broad-spectrum form of energy, when the period length is comparable to the phonon mean free path (MFP), there should always be a notable proportion of coherent phonon transport in superlattices. It is interesting to note that the 1/1 ML superlattices present a larger linear slope than that of the 2/2 ML ones. It implies a higher proportion of coherent phonons when the period length is reduced. Fig. 1d presents the temperature-dependent thermal conductivities of the 1/1 ML superlattices with varied period numbers. For the samples with 80, 120, and 160 periods (total thickness is about 48, 72, and 96 nm), the thermal conductivity increases as temperature increases and gradually reaches a plateau above 150 K. This behavior mainly comes from the boundary scattering since the thickness of the superlattice is comparable with the phonon MFP. When the period number further increases to 200 (about 120 nm), the Umklapp peak is observed around 150 K, indicating phonon-phonon inelastic scattering starts to contribute. We consider this single-crystalline-like behavior as another fingerprint of coherent phonon transport for which, to some extent, the interfaces in the superlattices are "invisible" to the traversing phonons. All these experimental findings corroborate the robust coherent phonon transport in the InAs/AlAs superlattices.

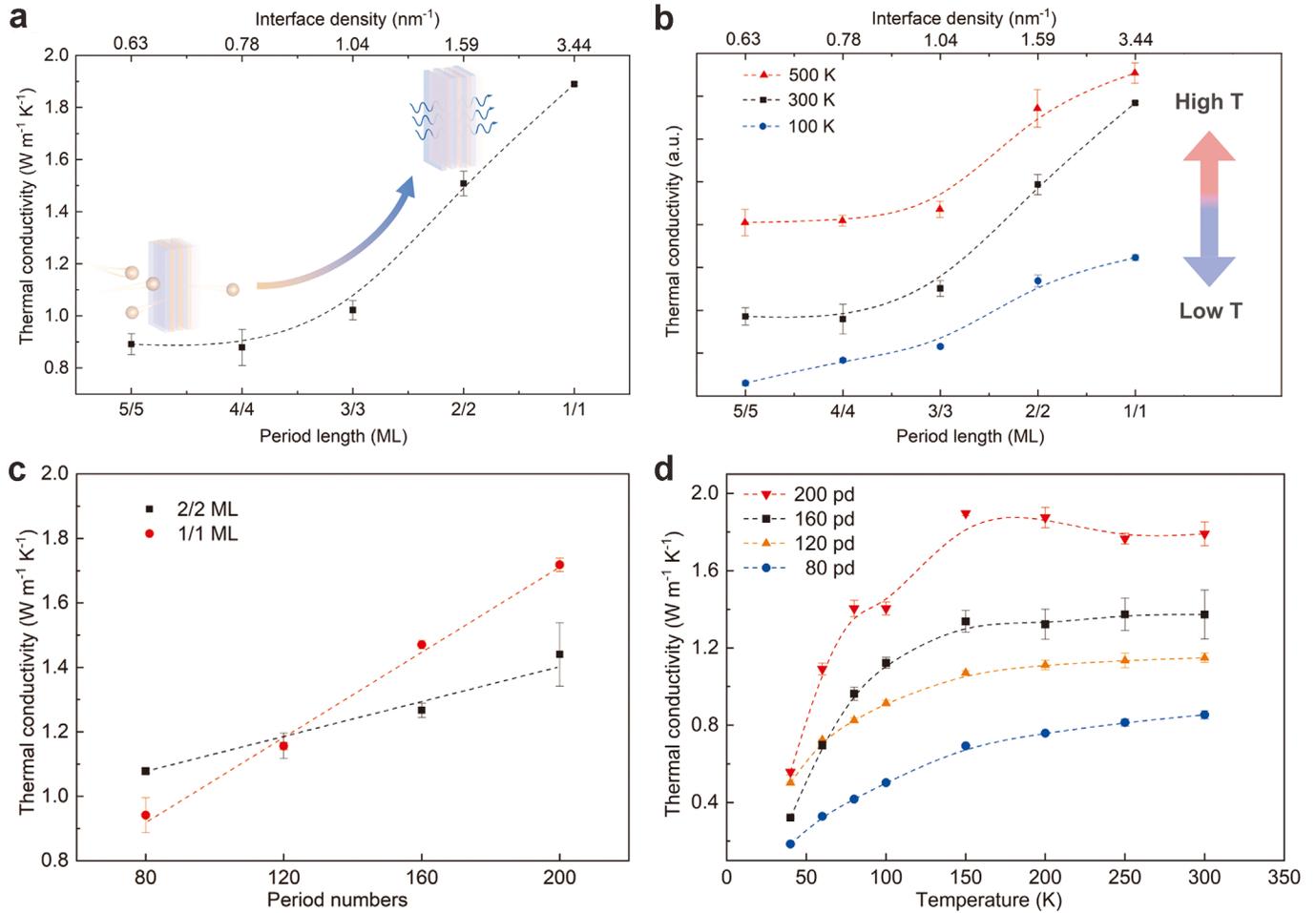

**Fig. 1. Thermal transport properties of InAs/AlAs superlattices. a**, Measured thermal conductivities of the superlattices as a function of period length along with the corresponding interface density at room temperature. The thermal conductivity of the InAlAs alloy is shown in Extended Data Fig. 4. The inset shows a schematic diagram of phonon transport mechanisms from a particle-like behavior to the wave-like one. **b**, Same as **a** but for various temperatures from 100 to 500 K. **c,** Measured room-temperature thermal conductivity as a function of the period number for the short-period (1/1 and 2/2 ML) superlattices. **d**, Measured temperature-dependent thermal conductivities for the 1/1 ML superlattices with different period (pd) numbers. The dashed lines are guides to the eye. The error bars represent the standard deviation of multiple measurements.

The observed robust coherent phonon transport is based on the high-quality epitaxial InAs/AlAs superlattices, which were characterized and proved by both XRD and TEM as shown in Fig. 2. Fig. 2a schematically illustrates the structural design of the InAs/AlAs superlattices. The XRD scans on (004) plane of these superlattices are plotted in Fig. 2b. For each sample, the XRD spectrum shows distinct satellite peaks owing to the superlattice periodicity along with a strong peak of the InP substrate. The shift of these satellite peaks provides direct evidence that the period lengths of the superlattices are successfully modulated to the designed values. Based on the XRD data, the actual interface densities of the superlattices can be calculated. The interfacial quality of the InAs/AlAs superlattices was evaluated by TEM. The cross-sectional high angle annular dark field (HAADF) STEM images of 2/2 and 5/5 ML superlattices, as representatives, are given in Figs. 2c and 2d, respectively. The images confirm good periodicity and distinct interfaces in these MBE-grown superlattices.

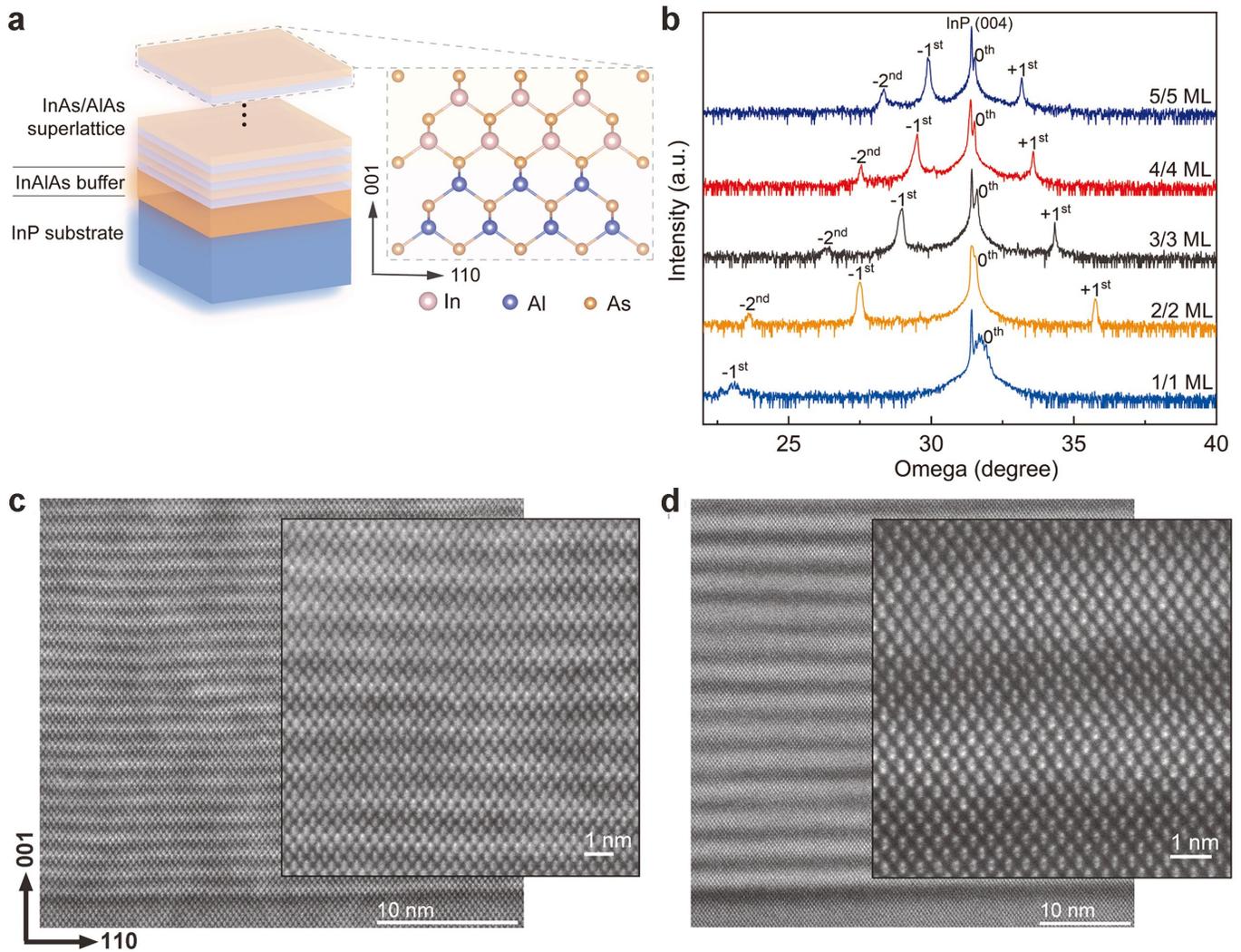

**Fig. 2. Structural design and characterizations of InAs/AlAs superlattices. a,** The schematic periodic structure of the superlattices. **b,** XRD omega-2theta scans of the superlattice samples with different period lengths, all showing the desired periodicity along the [001] direction. **c** and **d,** Cross-sectional HAADF-STEM images of the 2/2 and 5/5 ML superlattices, with different magnifications, both showing good periodicity along with distinct interfaces.

**Theoretical calculations and analyses**

To further understand the phonon transport mechanisms, we employed the first-principles calculations and Boltzmann transport equation with the relaxation time approximation to study the lattice thermal conductivities of the InAs/AlAs superlattices[43-46]. Figure 3a shows the calculated thermal conductivity for all the five superlattices at 300 and 500 K, respectively. The calculated values of thermal conductivities clearly show similar dependence on both period length and temperature as the experimental results[6]. Two evident trends can be drawn from the theoretical data. First, the thermal conductivity increases clearly with the decreasing period length at both temperatures. Second, for all the period lengths, the thermal conductivity at 500 K is lower than that at 300 K. This indicates enhanced Umklapp processes of phonons at a higher temperature. We calculated the thermal conductances of these superlattices based on the experimental data and the first-principles calculations (Extended Data Fig. 5). The thermal conductance shows evident increase when the layer thickness of the superlattices reduces down to ~1 nm, both at 300 and 500 K. The first-principles calculations agree quite well with the experimental results. In addition, this fashion of thermal conductance is in good agreement with Chen's theoretical study[25], demonstrating a phonon tunneling behavior that deriving from the wave nature of phonons. Our experiments prove that the phonon tunneling in InAs/AlAs superlattices is robust at relatively high temperatures up to 500 K. The period number dependent thermal conductances were also calculated for the 1/1 and 2/2 ML superlattices (Extended Data Fig. 6). For all period numbers, the 1/1 ML superlattices show significantly larger values (~twofold) in thermal conductance than that of the 2/2 ML ones. This finding agrees well with the aforementioned higher proportion of coherent phonons in the 1/1 ML superlattices. Regarding the period number dependence, the thermal conductances both decrease with increasing period numbers then tend to approach constants. Such behavior is resulted from the formation of mini bandgaps during the increasing period numbers, because the bandgaps suppress the phonon transport effectively[25].

To explore a deeper understanding of the coherent phonon transport mechanism in InAs/AlAs superlattices, we first analyzed the contribution of phonons of different frequencies to the thermal conductivity. It is found that low-frequency phonons below 4 THz contribute more than 90% to thermal conductivity (Extended Data Fig. 7). With longer-wavelength phonons dominating the thermal transport, robust phonon coherent transport and tunneling effect can be realized over a broad temperature range, even at higher temperatures. In this InAs/AlAs superlattice, the large mass contrast between In and Al plays an important role in "filtering" the high-frequency phonons[31].

Based on the first-principles calculations, more detailed information can be derived to understand the underlying mechanisms of phonon transport. Figure 3b shows the phonon MFP distribution as a function of phonon frequency for the 1/1 and 5/5 ML superlattices, respectively. The calculated MFP for other superlattices can be found in Extended Data Fig. 9. Clearly, low-frequency phonons present much longer MFPs, with maximum values reaching several microns. Such phonons can traverse coherently across the interfaces within the superlattices, and hence efficiently carry a major portion of heat through the structures. These results agree well with the finding discussed above, that the low-frequency phonons contribute dominantly to thermal conductivity. The phonon MFPs of the superlattices decrease with increasing period length in almost the entire frequency range. As shown in Fig. 3b, the MFPs of the 5/5 ML superlattice show an evident decrease compared with that of the 1/1 ML one. This phonon behavior is responsible for the specific trend of the thermal conductivity which increases with decreasing period length. Phonon MFP is the product of phonon group velocity $v$ and lifetime $\tau$. And it is considered an indicator to evaluate the scenario of interactions between phonons and the microstructures. To gain further insight into the phonon transport mechanisms, more information about the $v$ and $\tau$ is required. The lattice thermal conductivity $\kappa_L$ of a material can be simply written as a formula:

$$\kappa_L = \frac{1}{3} C \tau v^2 \qquad (1)$$

, where $C$ is the heat capacity. In the light of several milestone works on coherent phonon transport of superlattices, the interpretation is well accepted, that in the coherent regime, the decrease of thermal conductivity with increasing period length is a result of the reduced $v$ which is caused by the band-folding effect of phonons[6, 22, 36, 47]. However, our work indicates that the phonon lifetime also plays an important role in it and should be taken into account. We calculated and compared the phonon group velocities and lifetimes of InAs/AlAs superlattices with the 1/1 and 5/5 ML periods in Figs. 3c and 3d, respectively. Both the group velocity and lifetime show clear decrements with an increasing period length of the superlattices. These behaviors of phonons can be understood simply from the perspective of the band-folding effect. In the coherent regime, the band-folding effect can be tuned by modulating the period length of superlattices and it will affect the thermal transport from two aspects. First, band-folding increases the perturbation between phonon bands and thus decreases the phonon group velocity, especially for those high-frequency ones[22]. The calculated phonon dispersions of the 1/1 and 5/5 ML superlattices along the cross-plane direction (i.e. Γ-Z direction) are shown in Fig. 3e. The band-folding effect is revealed that the phonon dispersion of the 5/5 ML superlattice can be approximated as the folded one of the 1/1 ML superlattice. Then the phonon bands of the 5/5 ML superlattice become flatter due to the increased perturbation between each other. Flattened phonon bands yield lowered average phonon group velocities, as the velocity is described by the slope of the phonon bands: $v = \partial \omega_k / \partial k$, with $k$ the wave vector. Our results demonstrate that at lower frequencies (<4 THz) flat phonon bands can be achieved after folding, suggesting the successful modulation of long-wavelength phonons in this superlattice system. Second, phonon band-folding induces the formation of mini bandgaps, which can be clearly found at Γ and Z in the Brillouin zone for the 5/5 ML superlattice. With more mini bandgaps, the phonon modes are considered to be "squeezed" to a relatively smaller frequency space, which actually increases the phonon density of states. Such an effect results in an enhanced scattering between phonons, and hence shortens the phonon lifetime.

These calculations also reveal the mechanisms underlying the period-length and -number dependence of the thermal conductance of InAs/AlAs superlattices (Extended Data Fig. 5 and Fig. 6).

To quantitatively evaluate the impact of the band-folding effect on the phonon group velocity and lifetime, we calculated the averaged values over the entire Brillouin zone for the InAs/AlAs superlattices by employing the following equations[48]:

$$\bar{v}^\alpha = \sqrt{\frac{\sum_{\mathbf{q}s} C_V(\mathbf{q}s) v^\alpha(\mathbf{q}s) v^\alpha(\mathbf{q}s) \tau_{\mathbf{q}s}}{\sum_{\mathbf{q}s} C_V(\mathbf{q}s) \tau_{\mathbf{q}s}}}, \tag{2}$$

$$\bar{\tau}^\alpha = \frac{\sum_{\mathbf{q}s} C_V(\mathbf{q}s) v^\alpha(\mathbf{q}s) v^\alpha(\mathbf{q}s) \tau_{\mathbf{q}s}}{\sum_{\mathbf{q}s} C_V(\mathbf{q}s) v^\alpha(\mathbf{q}s) v^\alpha(\mathbf{q}s)}, \tag{3}$$

where $\mathbf{q}$ is the wave vector, $s$ is the dispersion branch, $C_V$ is the volumetric specific heat, and $\alpha$ denotes a particular direction in the Brillouin zone. Both the averaged phonon group velocity and lifetime along the cross-plane direction of the superlattices show an evident decrement with increasing period length. For example, the averaged group velocity decreases from 860.3 m/s for the 1/1 ML superlattice to 371.0 m/s for the 5/5 ML one, and accordingly, the averaged lifetime shortens from 44.3 to 14.7 ps. This simultaneous change in group velocity and lifetime implies the coexistence (or the duality) of the wave- and particle-like behaviors of phonons. In summary, our calculations reveal that by changing the period length of the superlattices, both the group velocity and lifetime can be tuned and responsible for the modulation of the phonon transport in these superlattices. These results well explain the experimental trend of thermal conductivity of the superlattices against period length.

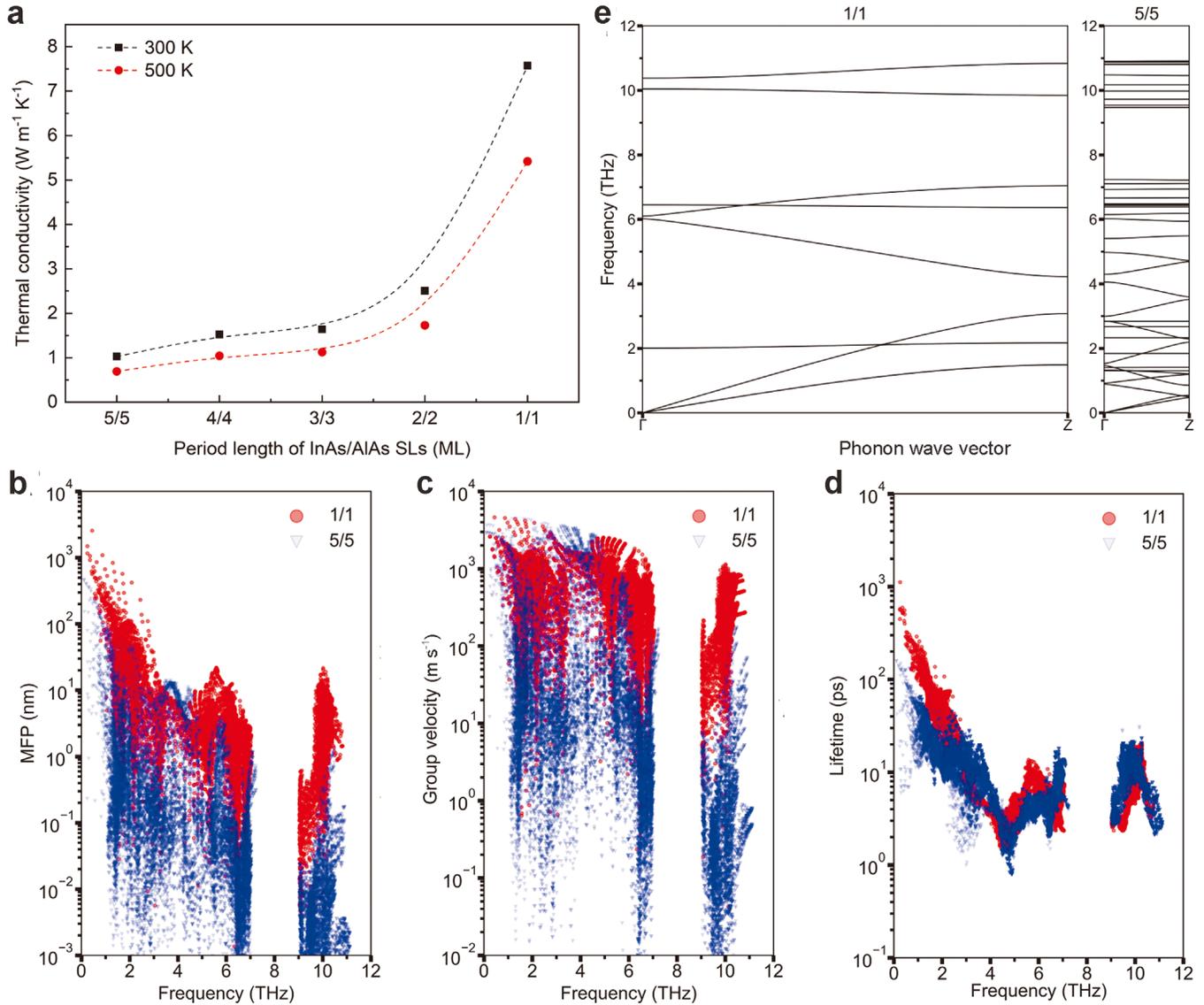

**Fig. 3. First-principles calculation results of InAs/AlAs superlattices. a**, Cross-plane lattice thermal conductivity as a function of period length at 300 and 500 K, respectively (the reason of the deviation between the calculated thermal conductivities and the experimental values is explained in Extended Data Fig. 8). Comparison of **b** phonon MFPs, **c** group velocity and **d** lifetime in the 1/1 and 5/5 ML superlattices as a function of frequency. **e,** Phonon dispersions of the 1/1 and 5/5 ML superlattices along the Γ-Z direction in the Brillouin zone, indicating a band-folding effect induced by period modulation. The phonon MFPs, group velocities, lifetimes, and dispersions of all the InAs/AlAs superlattices are given in Extended Data Fig. 9 and Fig. 10.

**Conclusions**

We have demonstrated solid experimental evidence of coherent phonon transport and tunneling effect in InAs/AlAs superlattices with strain compensation grown by MBE. The phonon coherence is robust and persists up to 500 K, attributed to the strategy that promotes the long-wavelength phonons in the transport. This high-temperature coherent phonon transport benefits from the high-quality superlattices, including the precise control of the period length in ML, highly ordered periodicity, high interfacial quality, etc. Our first-principles calculations corroborate the experiments along with the strategy and provide a comprehensive insight into the mechanisms underlying the phonon coherent behavior. Calculations show that not only phonon group velocity but also phonon lifetime can be reduced by band folding, which plays an important role in the thermal transport modulation. This also provides an evidence of the wave-particle duality of phonons. The mechanisms for the phonon coherence were extensively studied at different temperatures and proved robust coherence at elevated temperatures up to 500 K, which is the highest temperature reported. This observation may spark new ideas and strategies for thermal management in devices, especially those with complex superlattice structures, for their further developments towards even higher integration and power density. For example, the InAs/AlAs superlattices have shown potential in low-noise avalanche photodiodes[49, 50], and coherent phonons may offer opportunities to explore the underlying mechanism and optimize the device design.

**Extended Data Figure Legends:**

**Extended Data Fig. 1. Schematic setup of the TDTR system.**

**Extended Data Fig. 2. Data fitting of the room-temperature TDTR measurements for the InAs/AlAs superlattices with different period lengths from 1/1 to 5/5 ML.** The red open circles are experimental data points, the black solid line is the best fit from the model, and the blue dashed lines represent the best fits supposing the thermal conductivity is varied by +/-10%.

**Extended Data Fig. 3. Measured thermal conductivity of the InAs/AlAs superlattices as a function of period length at various temperatures.** The error bars represent the standard deviation of multiple measurements.

**Extended Data Fig. 4. Measured thermal conductivity of the InAs/AlAs superlattices as a function of period length at room temperature.** The data of the 50:50 InAlAs alloy sample (250 nm thick) is included as a comparison. The error bars represent the standard deviation of multiple measurements.

**Extended Data Fig. 5. Experimental and calculated thermal conductances of the InAs/AlAs superlattices as a function of period length along with the corresponding layer thickness, at 300 K (a) and 500 K (b) respectively.** The error bars represent the standard deviation of multiple measurements. 5% interfacial mixing is considered in the First-principles calculations. The theoretical data of a Si/Ge superlattice calculated with a transfer matrix method from ref. 25 are included in (a), and the data are tenfold reduced for ease of comparison. All these data share a similar fashion that the thermal conductance begins to increase when the layer thickness reduces to around 1 nm. This is a solid evidence of phonon tunneling which is derived from the wave nature of phonons. In addition, the First-principles calculations agree quite well with our experimental results.

**Extended Data Fig. 6. Experimental and calculated thermal conductances of the 1/1 and 2/2 ML InAs/AlAs superlattices as a function of period numbers at 300 K.** The error bars represent the standard

deviation of multiple measurements. 5% interfacial mixing is considered in the First-principles calculations. The calculations agree well with the experimental results. The thermal conductance decreases with increasing period numbers then tends to approach a constant.

**Extended Data Fig. 7. First-principles calculated contributions of phonons to the thermal conductivity as a function of phonon frequency for the InAs/AlAs superlattices with various period lengths ranging from 1/1 to 5/5 ML.** For all of these superlattices, the thermal conductivity is dominantly contributed by the low-frequency ($\leq 4$ THz) phonons.

**Extended Data Fig. 8. First-principles calculated thermal conductivity as a function of period length for the InAs/AlAs superlattices at 300 K in comparison with the experimental values.** The error bars represent the standard deviation of multiple measurements. Different mixing ratios of In and Al atoms in the interfacial layers were considered. When the mixing ratio comes to 5%, the calculated thermal conductivities agree well with the experimental values. Since interfacial mixing is an inevitable phenomenon that happens even for the most cutting-edge material growth technologies like MBE, it can be an important reason for the deviation between the calculation and experimental results. However, it should be reminded that in practice the interfacial imperfection is quite complicated and other defects will also cause a decrease in thermal conductivity.

**Extended Data Fig. 9. Calculated phonon mean free path (MFP) distribution (a) and group velocity distribution (b) as a function of phonon frequency for the InAs/AlAs superlattices with period length ranging from 1/1 to 5/5 ML.** The MFPs and the group velocities all show decrement with increasing period lengths of the superlattices.

**Extended Data Fig. 10. Calculated frequency dependent phonon lifetime distribution (a) and phonon dispersions (b) for the InAs/AlAs superlattices with period length ranging from 1/1 to 5/5 ML.** The

superlattices present shortened phonon lifetimes with increasing period lengths. The change in phonon dispersion indicates a band-folding effect induced by the period modulation.

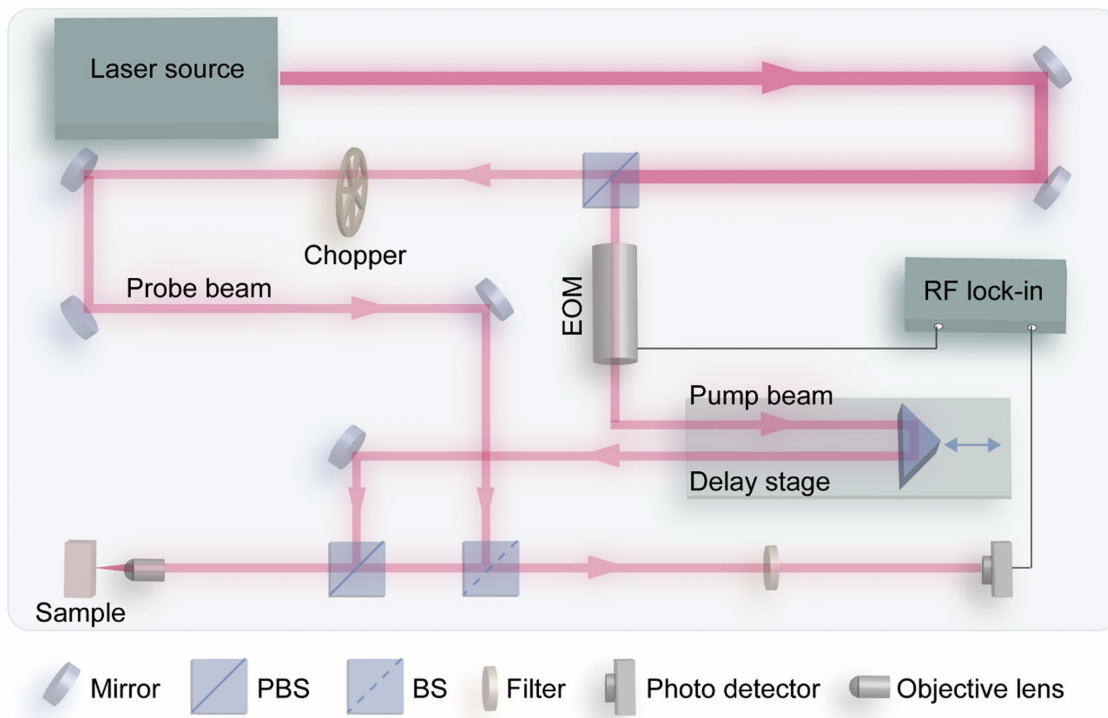

**Extended Data Fig. 1**

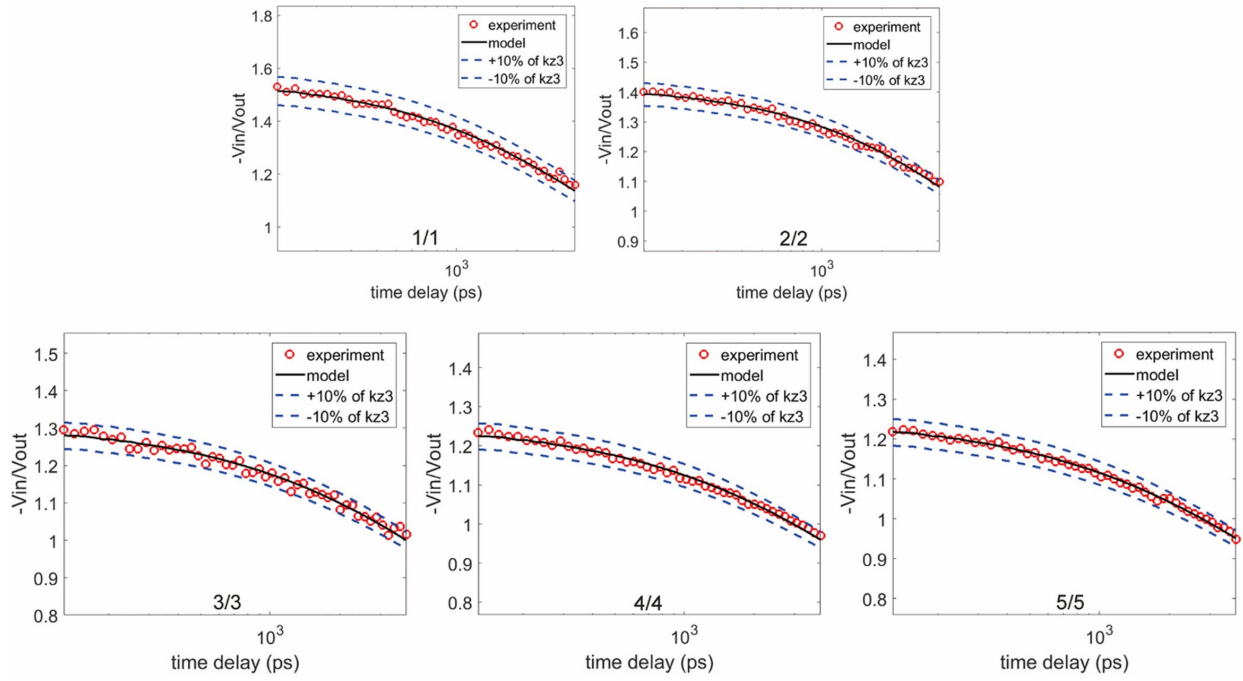

**Extended Data Fig. 2**

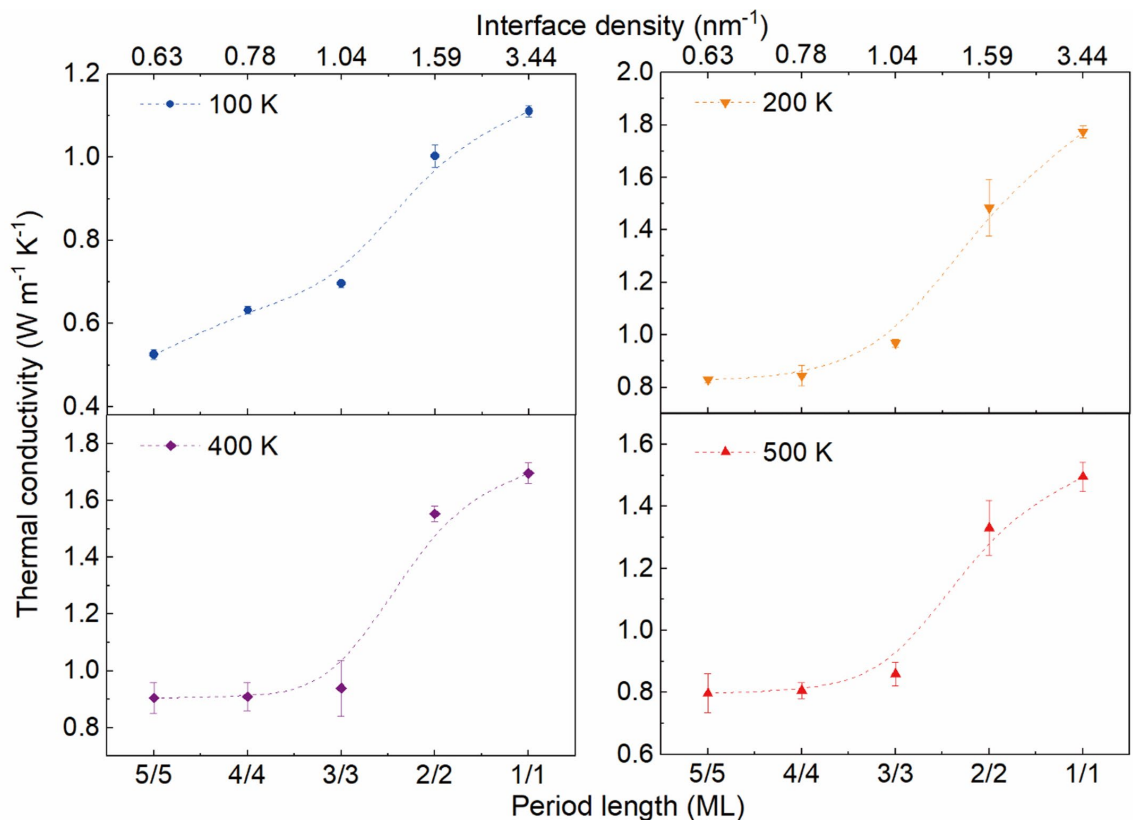

**Extended Data Fig. 3**

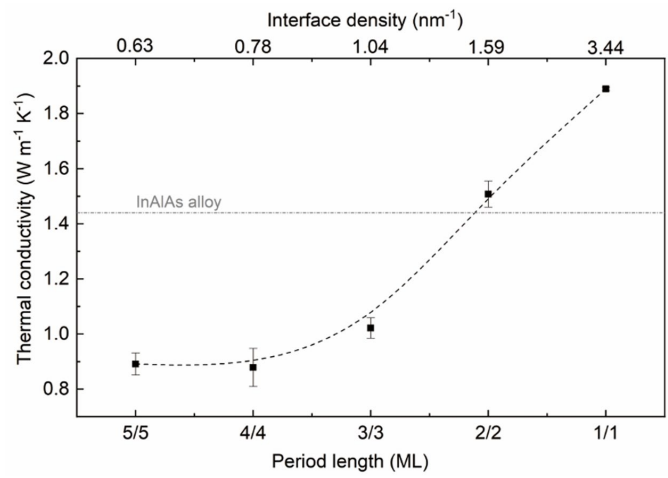

**Extended Data Fig. 4**

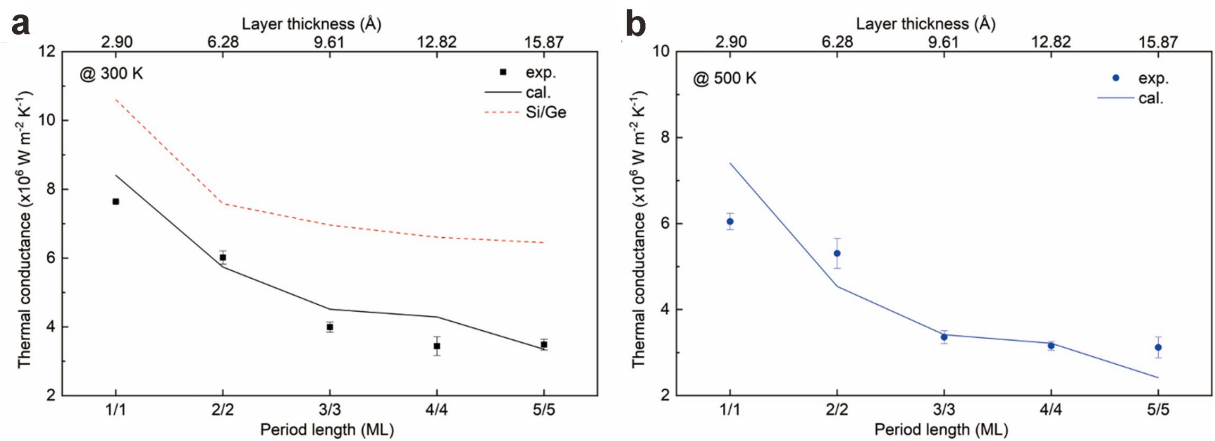

**Extended Data Fig. 5**

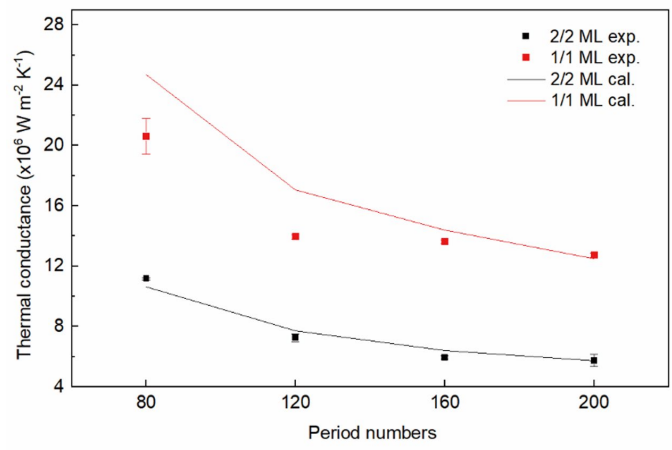

**Extended Data Fig. 6**

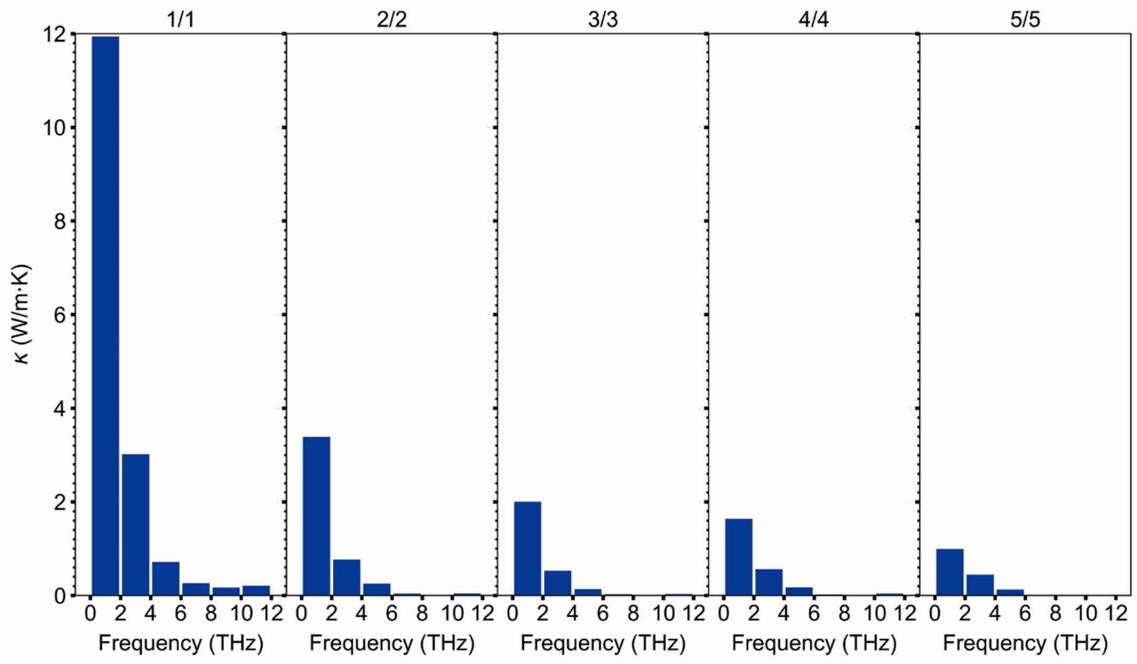

**Extended Data Fig. 7**

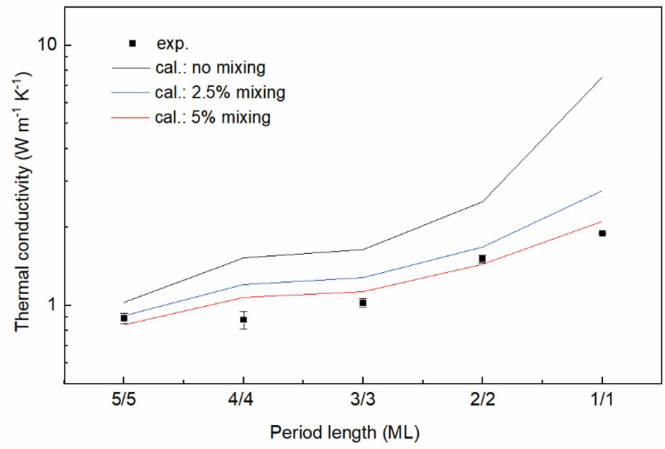

**Extended Data Fig. 8**

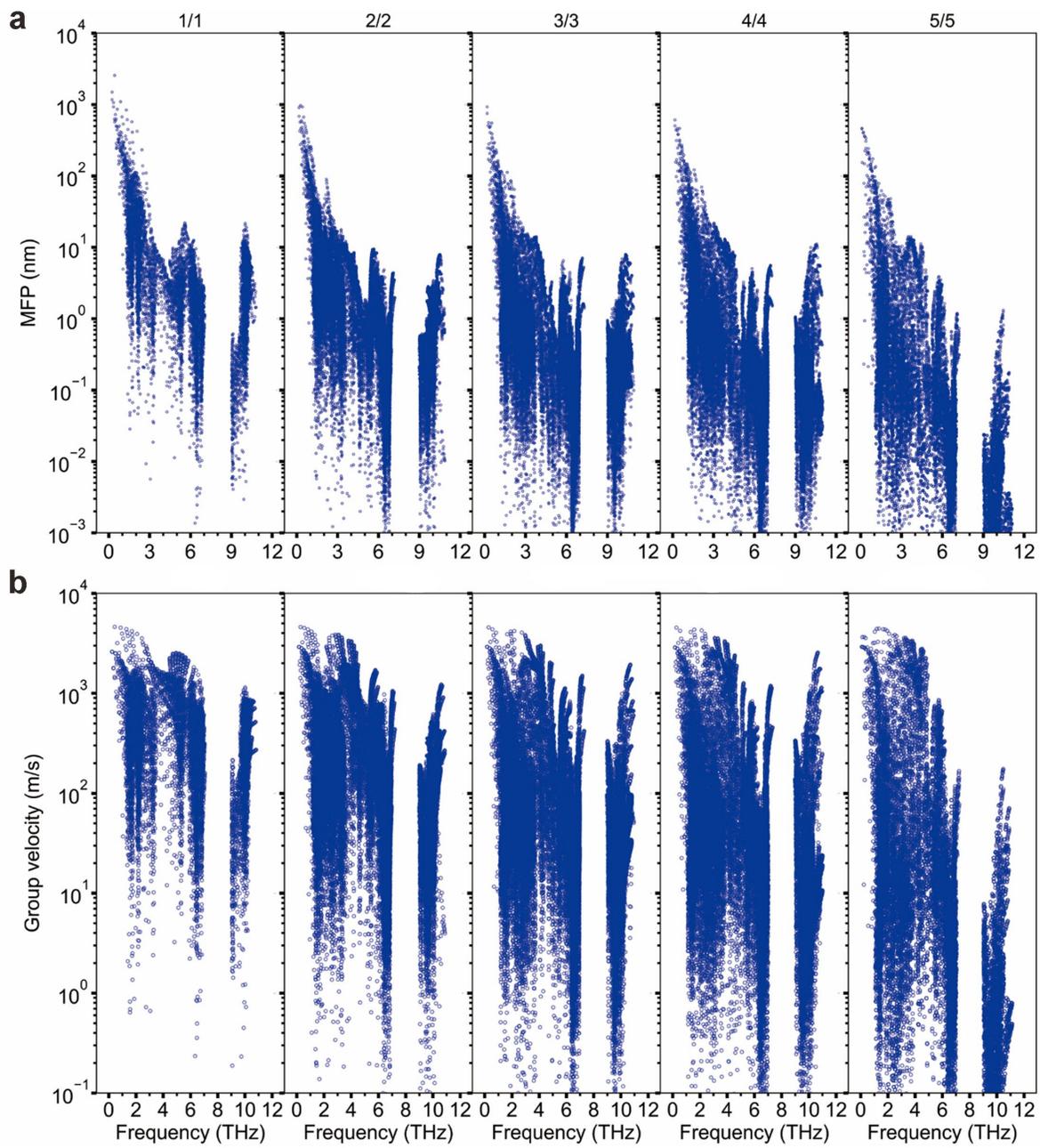

**Extended Data Fig. 9**

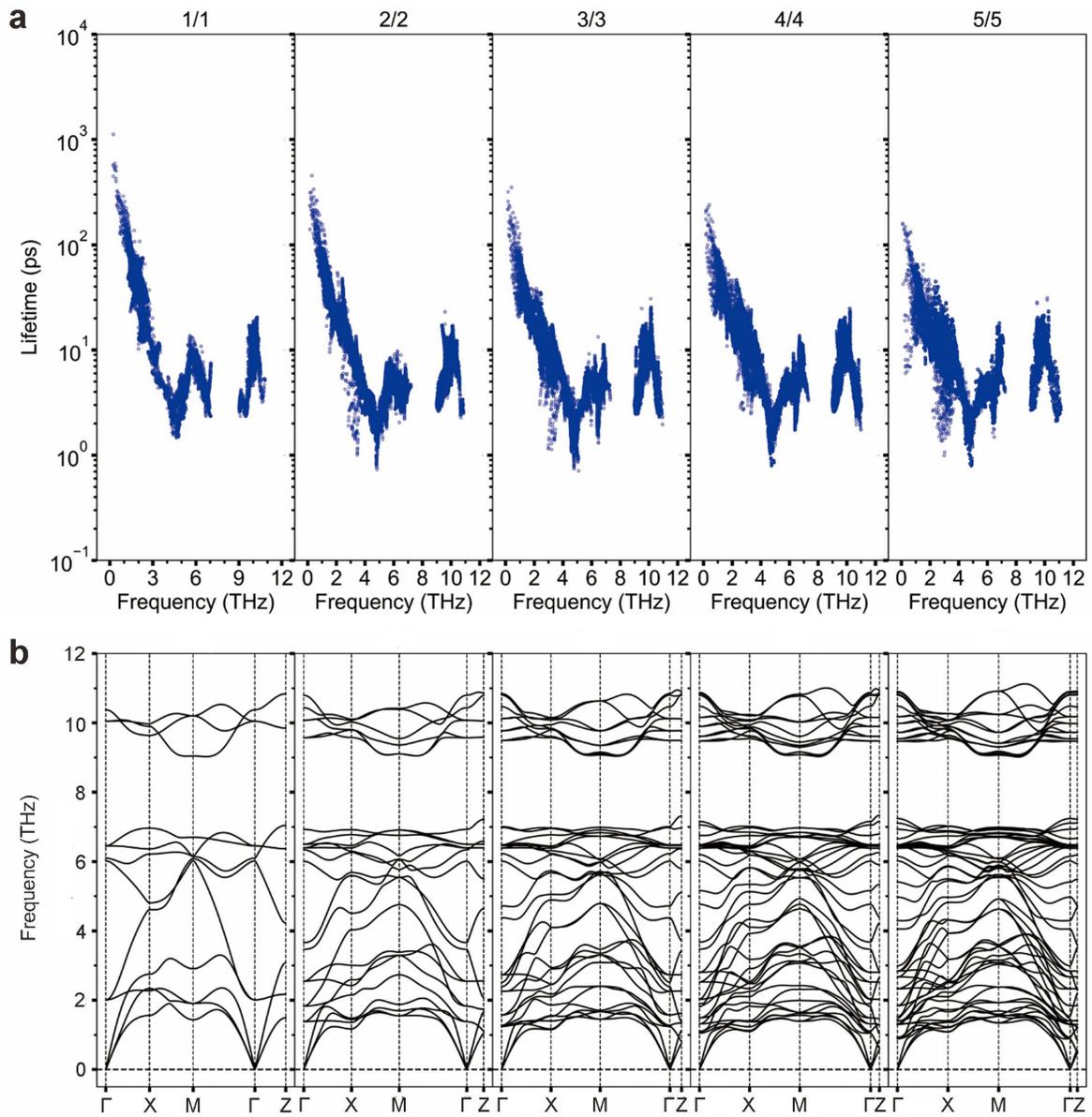

**Extended Data Fig. 10**